\documentclass[twocolumn,aps,prl,longbibliography]{revtex4-2}
\usepackage{amsmath}
\usepackage{amsfonts}
\usepackage{amssymb}
\usepackage[colorlinks=true,allcolors=blue]{hyperref}
\usepackage{graphicx}
\usepackage{xcolor}
\usepackage{mathrsfs}
\begin{document}

\title{Tailored high-contrast attosecond electron pulses for coherent excitation and scattering}
\author{Sergey V. Yalunin}
\email[]{yalunin@gwdg.de}
\author{Armin Feist}
\author{Claus Ropers}
\email[]{cropers@gwdg.de}
\affiliation{$^1$IV. Physical Institute, University of G\"ottingen, 37077 G\"ottingen, Germany}
\affiliation{$^2$Max Planck Institute for Biophysical Chemistry, 37077 G\"ottingen, Germany}

\begin{abstract}
Temporally shaping the density of electron beams using light forms the basis for a wide range of established and emerging technologies, including free-electron lasers and attosecond electron microscopy. The modulation depth of compressed electron pulses is a key figure of merit limiting applications. In this work, we present an approach for generating background-free attosecond electron pulse trains by sequential inelastic electron-light scattering. Harnessing quantum interference in the fractional Talbot effect, we suppress unwanted background density in electron compression by several orders of magnitude. Our results will greatly enhance applications of coherent electron-light scattering, such as stimulated cathodoluminescence and streaking.
\end{abstract}

\maketitle
Actively shaped free-electron beams offer the capability for enhanced sensing and microscopy \cite{Grillo2017a, Polman2019, Lourenco-Martins2021,Midgley2009, Kruit2016, Bliokh2017, GarciadeAbajo2021}, with numerous applications employing transverse shaping, incorporating tailored phase masks \cite{Uchida2010, Verbeeck2010, McMorran2011, Voloch-Bloch2013, Grillo2014, Grillo2017}, quasi-static electromagnetic optical elements \cite{McMorran2017, Harvey2017, Verbeeck2018, Pozzi2020} or scattering at the ponderomotive potential \cite{Kapitza1933, Freimund2001, Schwartz2019}. In the time domain, inelastic electron-light scattering (IELS) \cite{Barwick2009, GarciadeAbajo2010, Park2010,Feist2015,Kirchner2014} leads to correlated gain or loss of angular \cite{Cai2018, Vanacore2018} and linear momenta with energy \cite{GarciadeAbajo2016, Vanacore2019, Feist2020}. Following dispersive propagation, inelastic interactions can lead to a reshaping of the electron density into a train of attosecond pulses \cite{Sears2008, Hilbert2009, Baum2009, Feist2015}. These states have been experimentally prepared and characterized by quantum state tomography \cite{Priebe2017} or streaking \cite{Kozak2018, Morimoto2018, Schonenberger2019, Black2019}.

Besides their direct use for attosecond streaking and spectroscopy, such density-modulated electron beams are proposed for a multitude of applications, e.g., to imprint an external phase onto cathodoluminescence (CL) \cite{Pan2019a, Kfir2021, Karnieli2021, DiGiulio2021}, to induce a microscopic polarization in two-level systems (TLS), or to coherently build up mode amplitudes or local polarizations using independent electrons~\cite{Gover2020, Zhao2020, Ratzel2020, Reinhardt2020, Gover2020a, Morimoto2021}, thus promising a merger of electron microscopy with coherent spectroscopy.

A key limitation for these efforts is the quality of compression and the amount of uncompressed background density \cite{Priebe2017,Baum2017, Morimoto2018, Schonenberger2019, Reinhardt2020}, sometimes described by the classical multi-electron bunching factor for SASE-FELs \cite{Feng2014, Hirschmugl1991, Gover2019, Deng2021}. However, even for a pure single-electron state \cite{Bonifacio2009,Kling2015,Carmesin2020} focused in the quantum regime \cite{Priebe2017}, limited coherence arises \cite{Kfir2021,DiGiulio2021,Gover2020, Zhao2020, Gover2020a}.

In this Letter, we directly address this issue by devising an experimentally feasible scheme to prepare essentially background-free attosecond electron pulse trains, drastically enhancing the coherence in electron-light interactions. Introducing a single-color, multi-plane phase shaping approach, we predict tailored electron states representing excellent approximations to point-like classical currents within the optical cycle. These states are capable of producing CL with a near-unity degree of coherence, a maximized microscopic polarization of few-level systems, and the enhanced coherent build-up of mode and transition amplitudes using multiple independent electrons.

\begin{figure}
\center
\includegraphics[width=\columnwidth]{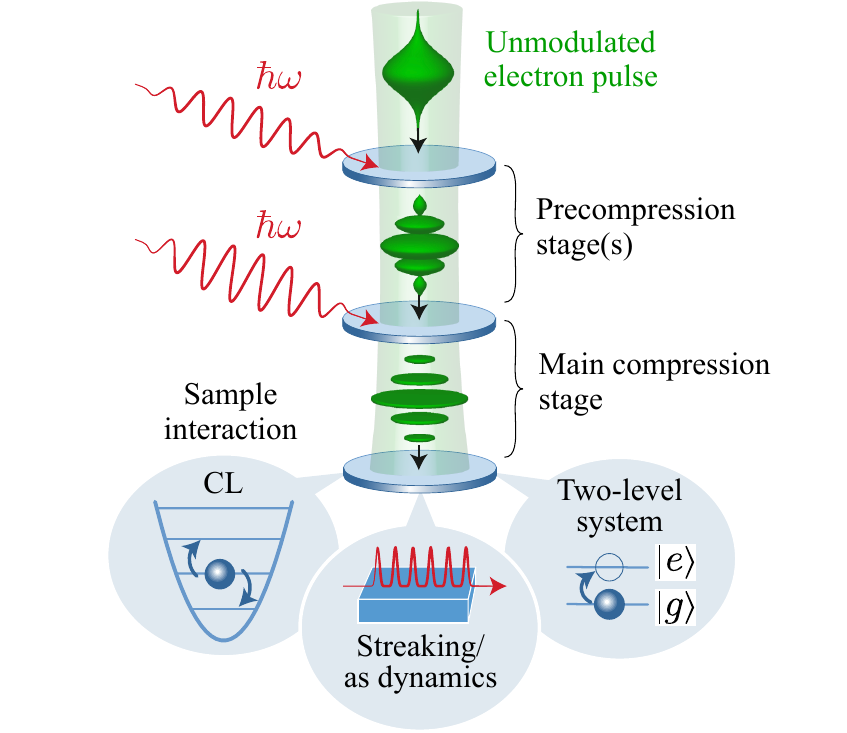}
\caption{Temporal electron pulse shaping and its prospective applications in attosecond physics, including streaking and the interaction with nanostructures and quantum systems.}\label{fig_sketch}
\end{figure}

The wavefunction shaping scheme is depicted in Fig.~\ref{fig_sketch}. It is based on sequential temporal phase plates using IELS to modulate the electron momentum, each followed by a drift stage to transfer this momentum modulation into a train of attosecond density pulses \cite{Priebe2017, Morimoto2018, Kozak2018}. As shown below, tailored precompression removes most of the electron density background and greatly enhances the coherence properties of a variety of excitations and scattering processes in the sample plane [Fig.~\ref{fig_sketch}, bottom].

The Letter is organized as follows. First, we briefly present a quantum mechanical approach to temporal phase plates and its classical limit. Second, the generation of nearly background-free attosecond electron pulses using sequential interactions is described. Finally, we discuss the consequences and features of using such electron states in selected applications, including streaking, CL and the excitation of two-level systems.

{\it Multislice method for electron propagation}.---There are various approaches to theoretically describe inelastic electron-light scattering and propagation~\cite{Barwick2009, GarciadeAbajo2010, Park2010, Feist2015}. Here, we employ a multislice approach~\cite{Cowley1957}, in which forward propagation of electrons along the $z$ axis is described by the effective Schr\"{o}dinger equation derived in the Supplemental Material~\footnote{See Supplemental Material for derivation of Eq.~(1), and simulated space-time evolution of the probability density after inelastic scattering}:
\begin{equation}\label{multislice}
i\hslash v\frac{\partial}{\partial z}\Psi(\mathbf{r},t)= \left[H_I(\mathbf{r},t)+D-{\mathcal E}\right] \Psi(\mathbf{r},t),
\end{equation}
where $v$ is the mean electron velocity, $\mathcal{E}=i\hslash \partial_t$ the translation operator, and $H_I(\mathbf{r},t)\approx -evA_z(\mathbf{r},t)$ the time-dependent scattering potential describing the interaction with light~\cite{Note1}. The dispersion operator $D$ is given by
\begin{equation}\label{eqD}
D = \frac{{\mathcal E}^2}{2\gamma^3 mv^2}-\frac{\hslash^2 }{2\gamma m} \nabla^2_{\perp},
\end{equation}
where $\gamma=1/\sqrt{1-v^2/c^2}$ is the Lorentz factor, $m$ the electron's rest mass, and $\nabla_{\perp}^2$ the Laplacian operator of the transverse coordinates. The Fresnel operator $U(h)=\exp(-iDh/\hslash v)$ describes the propagation of the electron state from $z$ to $z+h$ without interaction.
In the presence of an electromagnetic field, the solution of Eq.~(\ref{multislice}) can be obtained with the split-operator technique~\cite{Fleck1976}, using the Fresnel operator, the propagator for $D=0$~\cite{Park2010,Feist2020}, and one of the higher-order decomposition schemes~\cite{Yoshida1990,Suzuki1992}. Since we are mainly interested in temporal or longitudinal focusing, for simplicity, we assume that the interaction is independent of $x,y$, separating the wavefunction into its temporal $\psi(z,t-z/v)$ and transverse parts, where the latter is known analytically for cylindrical beams~\cite{McMorran2011,Lubk2018}. The advancement of the temporal part from $z$ to $z+h$ is obtained by the second-order expression:
\begin{equation}\label{split-operator}
\psi(z+h,t) = U(h/2) e^{i\Phi(z,t)}U(h/2)
\psi(z,t),
\end{equation}
where the phase function is given by
\begin{equation}\label{phase}
\Phi(z,t) = -\frac{1}{\hslash v}\int_z^{z+h}H_I(z,t+z/v) dz.
\end{equation}
This procedure is efficiently implemented with a fast Fourier transform algorithm. It is unitary and hence preserves the probability current $j=v|\psi(z,t)|^2$ integrated over time~\cite{Reimer2008}. It follows from Eq.~(\ref{split-operator}) that a scattering potential confined to an interval $(z,z+h)$ can be regarded as a thin inelastic or temporal phase plate located at $z+h/2$, by analogy with elastic phase plates~\cite{Schwartz2019,Konecna2020}. Thus, the forward propagation and temporal aberrations can be reduced to a phase function $\Phi(t)$ describing the temporal phase plate.

{\it Classical limit}.---We first outline the classical picture of temporal aberrations (${\hslash\to 0}$), where $\hslash\Phi(t)$ has the meaning of a classical action. Its time derivative defines the change of the electron's energy and velocity:
\begin{equation}\label{eq_u}
\Delta v(t) = -\frac{\hslash\Phi'(t)}{\gamma^3 m v}.
\end{equation}

Suppose that electrons uniformly distributed in time traverse a temporal phase plate at $z=0$ and gain a velocity change~(\ref{eq_u}). The change of the probability density with increasing $z$ is determined by the trajectories:
\begin{equation}\label{trajectory}
t(z) = t-\Delta v(t) z/v^2.
\end{equation}
The attracting fixed points in this map, i.e., the zeros of $\Delta v(t_0)$ with the time derivative $\Delta v'(t_0)>0$, correspond to paraxial temporal foci. Paths concentrate and form a caustic near such points, resulting in narrow peaks in the density~\cite{Baum2017}. For a time-harmonic phase $\Phi(t) = 2g\cos(\omega t)$ characterized by an effective interaction strength $g$ (see an expression in~\cite{Note1}) and a frequency $\omega$, the attracting points are given by $t_0=nT$, ${n\in \mathbb{Z}}$ with $T=2\pi/\omega$ being the optical period. The paraxial focus lies at the distance $l_f$ from the phase plate, with
\begin{equation}\label{l_Talbot}
l_f = \frac{v^2}{\Delta v'(t_0)} = \frac{l_T}{8\pi g}, \quad l_T= \frac{4\pi m\gamma^3 v^3}{\hslash\omega^2},
\end{equation}
where $l_T$ is the Talbot distance~\cite{DiGiulio2020}, which amounts to 200~mm for 120-keV electrons and 800-nm light.

\begin{figure}
\center
\includegraphics[width=\columnwidth]{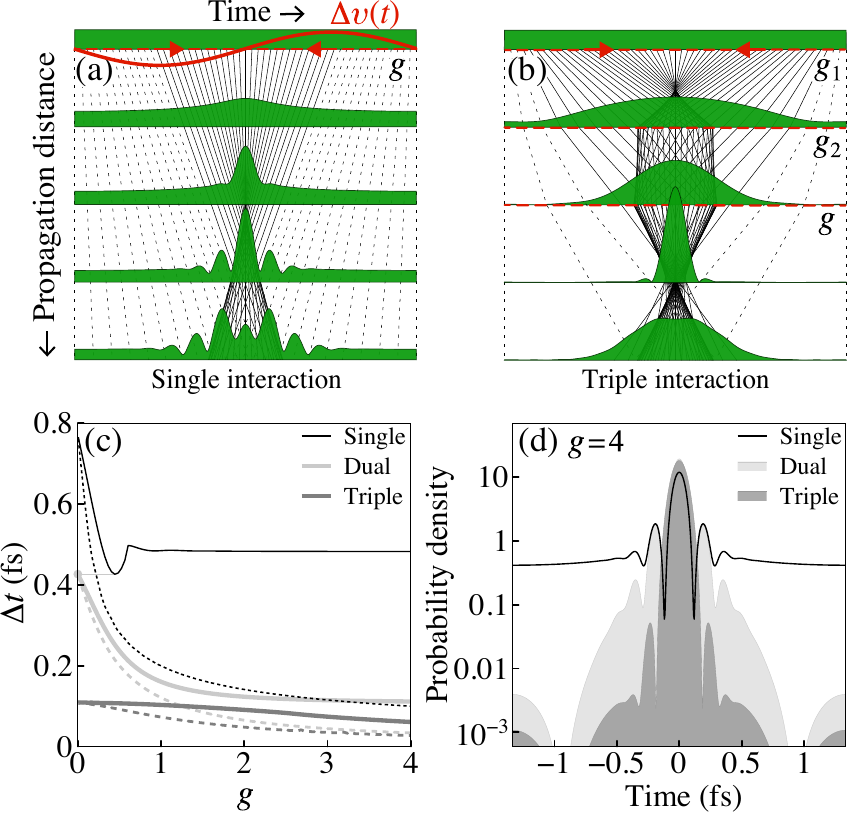}
\caption{(a),(b) Evolution of the magnitude of $\psi(z,t)$ within one optical period after single and triple interactions with 800-nm light. Calculation for 120-keV electrons and interaction strengths ${g_1=0.39}$, ${g_2=0.52}$, ${g=4}$. Solid and dotted black: Trajectories contributing in the central peak and background. (c) Optimized rms duration for single, dual and triple monochromatic interactions as a function of the main compression strength $g$ (and free variation of the precompression strengths). Dashed: Respective durations when the main compression is a parabolic phase modulation. (d) Semi-logarithmic plot of the density in the rms focus, illustrating a great reduction in the background density using sequential focusing.}\label{fig_space_time}
\end{figure}

Temporal focusing by monochromatic light is imperfect since paraxial trajectories do not converge into one point in the space-time diagram in Fig.~\ref{fig_space_time}(a), affecting the electron pulse duration. Another type of temporal aberration stems from repelling fixed points, at which half of the electrons are steered to a nearly homogeneous background in density [Fig.~\ref{fig_space_time}(a)]. The latter does not improve with increasing $g$ because it depends only on the product $gz$ (see Eq.~(\ref{trajectory})), and also phase-squeezed light does not reduce the background~\cite{DiGiulio2020}. In principle, both types of temporal aberrations could be eliminated using multiple harmonics to approach a parabolic phase modulation of the form $\Phi(t) = -g\omega^2s^2(t)$, where ${s(t)=(t+T/2 \ \textrm{mod}\ T)-T/2}$ is the periodic saw-like function with zeros $nT$, $n\in \mathbb{Z}$. Generalized electron beam shaping using multiple harmonics has recently been theoretically considered \cite{Reinhardt2020}, and two-color phase modulation experiments have been performed in the context of attosecond focusing and quantum state reconstruction \cite{Priebe2017}. However, superimposing an even larger number of harmonics with controlled amplitude and phase may render this approach rather impractical for attosecond focusing. Instead, as we show in the following, sequential monochromatic interactions in separate planes represent an even more powerful and experimentally tractable scheme to address temporal aberrations.

{\it Eliminating focusing aberrations by sequential scattering}.---As described above, the aberrations in temporal focusing are most severe for those parts of the probability density furthest from the center of the cycle. We show here that weak precompression stages of interaction strength less than unity can reshape the density for much more efficient focusing [cf. Fig.~\ref{fig_space_time}(b)]. Specifically, the fractional Talbot effect \cite{Berry1996} offers a powerful way to suppress the density near the repelling points, without adding further phase aberrations. At one fourth of the Talbot distance, the wavefunction around the attractive points is a superposition of the phase-modulated initial state $\psi(0,t)\approx\psi_0(t_0)\exp[2ig_1\cos(\omega t)]$ and its replica shifted by half of the optical period~\cite{Guigay1971,Cloetens1997}:
\begin{equation}\label{two_pulses}
\psi(l_T/4,t)=\frac{\exp(-i\pi/4)}{\sqrt{2}}\left[\psi(0,t)+i\psi(0,t+T/2)\right].
\end{equation}
Importantly, due to destructive interference, the probability density
\begin{equation}
\rho(l_T/4,t)\approx|\psi_0(t_0)|^2\{1+\sin\left[4g_1\cos(\omega t)\right]\}
\end{equation}
vanishes at the repelling points $t=t_0\pm T/2$ when $g_1=\pi/8$. While such a modulation itself does not provide much temporal compression, a second stronger temporal phase plate completes the focusing, but now with greatly reduced background density.

\begin{figure}
\center
\includegraphics[width=\columnwidth]{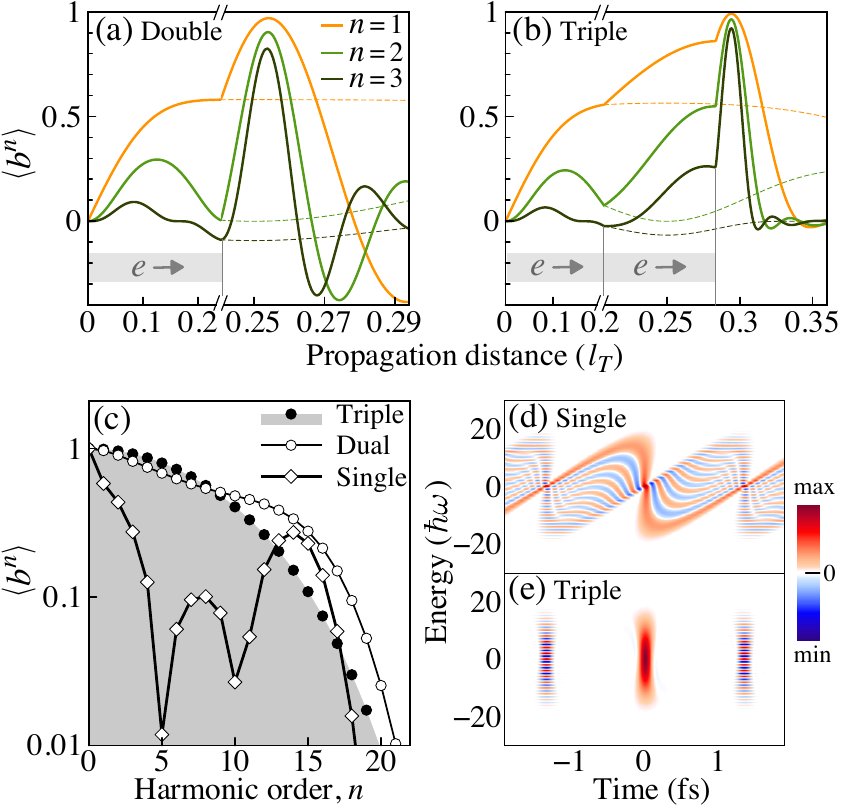}
\caption{(a),(b) Evolution of $\langle b^n\rangle$ (solid lines) after dual (${g_1=0.44}$) and triple (${g_1=0.39}$, ${g_2=0.52}$) interactions with the compression strength ${g=4}$. Dashed: The evolution after the first interaction only. (c) Values of $\langle b^n\rangle$ obtained at the distance where $\langle b\rangle$ reaches a maximum. (d),(e) Wigner distributions in the focus, showing a localization of electrons in phase space.}\label{fig_coherence}
\end{figure}

Depending on the specific application, several properties can be considered for assessing the quality of attosecond focusing. A good measure of background density is given by the root-mean-squared (rms) duration $\Delta t(z)$,
\begin{equation}\label{delta_t}
\Delta t(z)=\sqrt{\langle s^2\rangle},\quad \langle s^2\rangle=\frac{1}{C}\int_{-\infty}^{\infty}s^2(t)\rho(z,t)dt,
\end{equation}
where $\rho(z,t)=|\psi(z,t)|^2$ is the probability density, $C$ its integral, and $s(t)$ the saw-like function. Figure~\ref{fig_space_time}(c) illustrates the significant improvement of attosecond focusing in terms of the rms duration upon adding one or two prefocusing stages. As a function of the main compression strength $g$, $\Delta t(z)$ is evaluated at the respective minima along $z$, for a single interaction, a parabolic phase profile, and the optimized dual and triple (two weak and one strong) interactions. It is evident that the single-interaction case is rather limited in terms of its achievable rms duration, being minimized near $g\approx0.45$ at $\Delta t\approx 0.426$~fs. By an effective suppression of background density [see Fig.~\ref{fig_space_time}(d)], the dual interaction substantially reduces $\Delta t$. Although the most substantial absolute improvement is already obtained for the case of a single precompression phase plate, adding additional interaction planes leads to a further optimization, as evident from the results for a total of three phase plates shown in Figs.~\ref{fig_space_time}(b)-\ref{fig_space_time}(d) (see also Ref.~\cite{Note1}). It is worth noting that the idealized parabolic phase modulation applied after the precompression stages leads to only minor additional improvements [see Fig.~\ref{fig_space_time}(c)], illustrating that the introduced prefocusing scheme indeed strongly reduces aberrations and provides rms durations very close to the theoretical minimum.

{\it Applications of background-free pulses}.---We note that the background-free attosecond pulses produced by this scheme will be exceedingly useful in a number of recently proposed schemes for coherent interactions of free electrons with nanostructures and individual quantum systems~\cite{DiGiulio2019,BenHayun2021}. This includes electron-mediated coherence transfer and the coherent build-up of excitations for multiple subsequent electrons~\cite{Kfir2021,Pan2019a,DiGiulio2021, Gover2020, Zhao2020, Ratzel2020, Reinhardt2020, Gover2020a, Morimoto2021}. In each of the underlying processes, the modulation amplitude of the electron density at the fundamental modulation frequency, or its harmonics, plays a major role. This is quantified by the expectation values or moments  $\langle b^n\rangle$ of the ladder operator $b=\exp(i\omega t)$~\cite{Feist2017, Zhao2020}. Different terminologies relating to these expectation values have been used, including the degree of coherence \cite{Kfir2021}, the coherence factor~\cite{DiGiulio2020}, or the bunching factor~\cite{Schroeder2001, Gover2019, Pan2019a, Deng2021}.

\begin{figure}
\center
\includegraphics[width=\columnwidth]{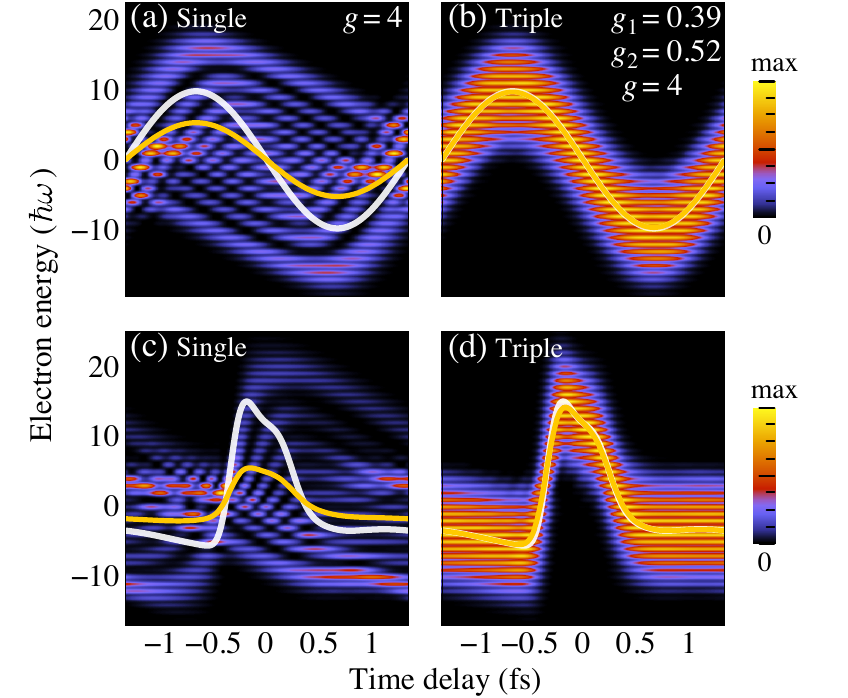}
\caption{Field-driven streaking spectrograms simulated with the electron pulses shown in Fig.~\ref{fig_space_time}(d). White lines: Temporal profile of the streak field $-\hslash\Phi'(\tau)$, where $\Phi(\tau)$ is given by Eq.~(\ref{phase}). Yellow: Calculated mean electron energy, $E(\tau)$.}\label{fig_streaking}
\end{figure}

The moments arising from a single interaction, $\langle b^n\rangle=J_n[4g\sin(2\pi n z/l_T)]$ (see also Ref.~\cite{Zhao2020}), are limited by the maxima of the Bessel functions $J_n$. Our sequential interaction scheme results in a substantial and nearly simultaneous enhancement of $\langle b^n\rangle$ above these values, as shown in Figures~\ref{fig_coherence}(a)-\ref{fig_coherence}(c). Here, we numerically optimized the first moment $\langle b\rangle$ by a variation of the precompression strengths $g_{1,2}$ and the respective distances $d_{1,2}$ between interaction planes, for a fixed value of $g=4$. For a dual interaction, the optimization yields a maximum value of $\langle b\rangle_{max}\approx 0.97$ for $g_1\approx 0.44$ and $d_1\approx 0.24 l_T$, well above the theoretical maximum $\langle b\rangle_{max}=\max[J_1(z)]\approx 0.58$ for a single interaction, discussed in Ref.~\cite{DiGiulio2021,Zhao2020}. Note that the optimum distance $d_1$ is indeed very close to the quarter Talbot distance. An additional precompression phase plate $g_2$ leads to a further enhancement [Fig.~\ref{fig_coherence}(b)], with values of $\langle b\rangle_{max}\approx 0.99$ and 0.998 for $g=4$ and $20$, respectively. The enhancements of the higher moments are equally striking, reaching a nearly Gaussian distribution spanning multiple orders for a triple interaction [Fig.~\ref{fig_coherence}(c)].

The background-free attosecond focusing has direct implications for electron-driven radiative emission in the form of CL \cite{Bendana2011, Lin2018, Mignuzzi2018, Talebi2019_merging, Schilder2020, vanNielen2020, Christopher2020, Remez2019}. Specifically, our results imply a degree-of-coherence of CL near unity \cite{Pan2019a, Kfir2021, Karnieli2021, DiGiulio2021}. Similarly, due to the pronounced localization in time [see Fig.~\ref{fig_space_time}(d)] and the large moments of $b$, the background-free attosecond electron pulses lead to an almost fully coherent excitation of TLSs with a transition energy $\Delta E=\hslash\omega$~\cite{Gover2020, Zhao2020, Ratzel2020, Reinhardt2020, Gover2020a, Morimoto2021}. The quantum state in this case can be readily described using the Bloch model of the density matrix~\cite{Scully1997}. For a TLS initially in the ground state, the excitation can be seen as a ``rotation'' of the Bloch vector $\mathbf{a}=-\hat{\bf z}$ around a unit vector $\mathbf{n}$ in the $x$-$y$ plane, $\mathbf{a}(\theta)=\mathbf{a}\cos\theta+\langle b\rangle (\mathbf{n}\times \mathbf{a})\sin\theta$, where $\theta$ is the rotation angle. This angle can be determined through the transition probability $P_2=\sin^2(\theta/2)$ and Eq.~(11) in Ref.~\cite{Gover2020}. Unmodulated electrons, or electrons with small $\langle b\rangle$, lead to a loss of the partial information about the relative phase between the eigenstates of the TLS, because of the large uncertainty in the transition time~\cite{GarciadeAbajo2021a}. In contrast, modulated electrons with $\langle b\rangle\approx 1$ retain this information and lead to a purity of the final state of $\textrm{Tr}(\rho^2)=1+\frac{1}{2}(\langle b\rangle^2-1)\sin^2\theta$ near unity. Incidentally, the dipole moment of the final state is maximized. Moreover, due to minimized quantum entanglement between the interacting systems, subsequent excitations driven by independent electrons modulated by the same reference wave can coherently build up either the cathodoluminescence in a particular mode or the transition amplitude in a TLS~\cite{Gover2020, Zhao2020,GarciadeAbajo2021}, leading to transition probabilities scaling with $N^2$ for small $\theta$, where $N$ is the number of electrons incident within the decoherence or dephasing time.

We note that the electron pulses produced here can be regarded as optimized approximations to point-like, classical phase-space densities, as is clearly seen by their Wigner function shown in Fig.~\ref{fig_coherence}(e). This makes them ideally suited for temporal probing of periodic electromagnetic fields by means of free-electron energy or angular streaking~\cite{Kirchner2014,Kozak2018, Morimoto2018, Schonenberger2019,Black2019}. In energy streaking, a time-periodic electric field $F(\tau)=\sum_n c_ne^{in\omega \tau}$, or more precisely its spatial Fourier components in Eq.~(\ref{phase}), are mapped onto the energy domain via the expression: $E(\tau)\sim\langle F(t+\tau)\rangle = \sum_n \langle b^n\rangle c_ne^{in\omega \tau}$,
where $E(\tau)$ is the displacement of the mean energy of electrons, and $\tau$ the time delay between the electron pulse train and the streak field. Of course, the rather limited moments $\langle b^n\rangle$ resulting from single-field focusing [see Fig.~\ref{fig_coherence}(c)] strongly affect the performance of the streaking technique. Furthermore, as shown in Figs.~\ref{fig_streaking}(a) and \ref{fig_streaking}(c), the presence of a background in electron pulses produced with the conventional scheme leads to pronounced spectral distortions. In contrast, the background-free electron pulses produced with our sequential focusing scheme lead to an almost perfect temporal representation of the streak field [see Figs.~\ref{fig_streaking}(b) and \ref{fig_streaking}(d)].

Possible experimental implementations of our focusing scheme can be realized over a wide range of energies accessible in scanning and transmission electron microscopes. The main design parameter to consider is the Talbot distance [Eq.~(\ref{l_Talbot})], which scales with the cube of the electron velocity. Specifically, the pre-compression distance $l_T/4$ at 800~nm optical wavelength reduces from 120~mm at 200 keV to 8.6~mm at 40~keV and 1~mm at 10~keV, dimensions for which technical solutions should readily be found.

{\it Conclusion}.---In summary, we have theoretically demonstrated how the concept of sequential scattering by light and the fractional Talbot effect can be applied to generate high-contrast attosecond electron pulses. We achieve an efficient correction of temporal aberrations, which leads to greatly enhanced coherence properties in electron-light scattering as compared to conventional temporal compression. Although the core of our generation scheme is a true quantum effect, harnessing destructive interference of the wavefunction, the resulting electron states exhibit an almost entirely positive Wigner function and greatly improved localization in phase space. Such states closely represent classical currents and produce practically fully coherent excitation and radiation at the modulation frequency. Finally, we believe that similar schemes may be applicable also for free-electron lasers operating in the quantum regime \cite{Bonifacio2009, Kling2015, Carmesin2020}.

\begin{acknowledgments}
We thank K.\,E.\,Priebe, Th.\,Rittmann, H.\,Louren\c{c}o-Martin, O.\,Kfir, V.\,Di\,Giulio and F.\,J.\,Garc\'{i}a\,de\,Abajo for insightful discussions. This work was funded by the Deutsche Forschungsgemeinschaft (DFG, German Research Foundation) - 432680300/SFB\,1456 (project C01), 255652344/SPP\,1840 (project ‘Koh\"arente Wechselwirkungen starker optischer Nahfelder mit freien Elektronen’), the Gottfried Wilhelm Leibniz program, and the European Union’s Horizon 2020 research and innovation programme under grant agreement No 101017720 (FET-Proactive EBEAM).
\end{acknowledgments}
\bibliography{references}
\end{document}


\title{Tailored high-contrast attosecond electron pulses for coherent excitation and scattering}
\author{Sergey V. Yalunin$^1$}
\author{Armin Feist$^1$}
\author{Claus Ropers$^{1,2}$}
\affiliation{$^1$IV. Physical Institute, University of G\"ottingen, 37077 G\"ottingen, Germany}
\affiliation{$^2$Max Planck Institute for Biophysical Chemistry, 37077 G\"ottingen, Germany}

\maketitle
\section{Derivation of Eq.~(1)}
The relativistically corrected Schr\"odinger equation has been previously obtained in Refs.~\cite{Friedman1988,Gover2018}. Here we provide a derivation of another form of the relativistic Schr\"odinger equation which, in contrast to the conventional Schr\"odinger equation, is linear in the derivative $\partial_z$ and quadratic in $\partial_t$. The quadratic term describes dispersion of the electron wavefunction in time. This form is particularly useful for numerical computations by means of the multislice method~\cite{Kirkland2010} and fast Fourier transform algorithms.

As in Ref.~\cite{Friedman1988}, our starting point is the Klein-Gordon equation (in SI units):
\begin{equation}\label{KG}
\mathcal{E}^2\Psi=[c^2(\mathbf{p}-e\mathbf{A})^2+m^2c^4]\Psi,
\end{equation}
where $e=-|e|$ is the electron's charge, and $\mathbf{A}$ the vector potential in the Coulomb gauge ($\nabla\cdot\mathbf{A}=0$). We omitted the scalar potential, since it is zero in the absence of external charges. Removing the central energy and momentum of electrons by the substitution $\Psi\to\Psi \exp[i\gamma m (\mathbf{v}\cdot\mathbf{r}-c^2t)/\hslash]$, we rewrite Eq.~(\ref{KG}) in an alternative, but equivalent form:
\begin{equation}\label{eq1}
\mathbf{v}\cdot(\mathbf{p}-e\mathbf{A})\Psi(\mathbf{r},t)= \left[\mathcal{E}+\frac{\mathcal{E}^2}{2\gamma mc^2}-\frac{(\mathbf{p}-e\mathbf{A})^2}{2\gamma m}\right]\Psi(\mathbf{r},t),
\end{equation}
where $\mathbf{v}\cdot\mathbf{p}=-i\hslash v\partial_z$, since $\mathbf{v}$ is chosen along the $z$ axis.

We make the common assumption of multislice methods that the spread in kinetic momentum is small compared to the central momentum of the electrons $p_0=\gamma mv$. This is justified, as transmission electron microscopes operate at 30-300~kV acceleration voltages, while typical energy spreads in the experiments are of the order of 1-300~eV. This ensures that the second and third terms in the square brackets in Eq.~(\ref{eq1}) are small compared to $\mathcal{E}$. The third term can be approximated by $(\mathbf{p}-e\mathbf{A})^2/2\gamma m \approx (\mathbf{p}-e\mathbf{A})_{\perp}^2/2\gamma m+\mathcal{E}^2/2\gamma mv^2$ and combined with the second term to give $-\mathcal{E}^2/2\gamma^3 mv^2-(\mathbf{p}-e\mathbf{A})_{\perp}^2/2\gamma m$. This leads us to Eqs.~(1) and (2) in the main text, and the following representation for the scattering potential:
\begin{equation}\label{eqHI}
H_I(\mathbf{r},t)=-evA_z(\mathbf{r},t)+\frac{e^2\mathbf{A}_{\perp}^2(\mathbf{r},t)}{2\gamma m}-\frac{e\mathbf{p}_{\perp}\cdot\mathbf{A}_{\perp}(\mathbf{r},t)}{\gamma m}.
\end{equation}
We note that the second (quadratic) term on the right-hand side of this equation is important only at very strong electromagnetic fields. The third term is also small compared to the first term, especially for highly collimated electron beams.

\section{Phase function $\Phi(t)$}
Consider a derivation of the time-harmonic phase modulation which is the key element of our temporal focusing scheme. It follows from Eq.~(4) (in the main text) that the phase function $\Phi(t)$ depends on the temporal form of the scattering potential~(\ref{eqHI}). We consider a time-harmionic field characterized by the vector potential $A_z(\mathbf{r},t)=\textrm{Re}\{E_{0z}(z) e^{-i\omega t}\}/\omega$, where $E_{0z}(z)$ is the complex electric amplitude. Retaining only the first term in Eq.~(\ref{eqHI}), we arrive at the following expression for the phase function:
\begin{equation}
\Phi(t)=2\textrm{Re}\{ge^{-i\omega t}\}=2|g|\cos[\omega t-\arg(g)],
\end{equation}
where
\begin{equation}
g = \frac{e}{2\hslash\omega} \int_{-\infty}^{\infty} E_{0z}(z)e^{-i\omega z/v} dz.
\end{equation}

\section{Supplement to Fig.~2}
\begin{center}
\begin{figure}[h]
\includegraphics[width=\columnwidth]{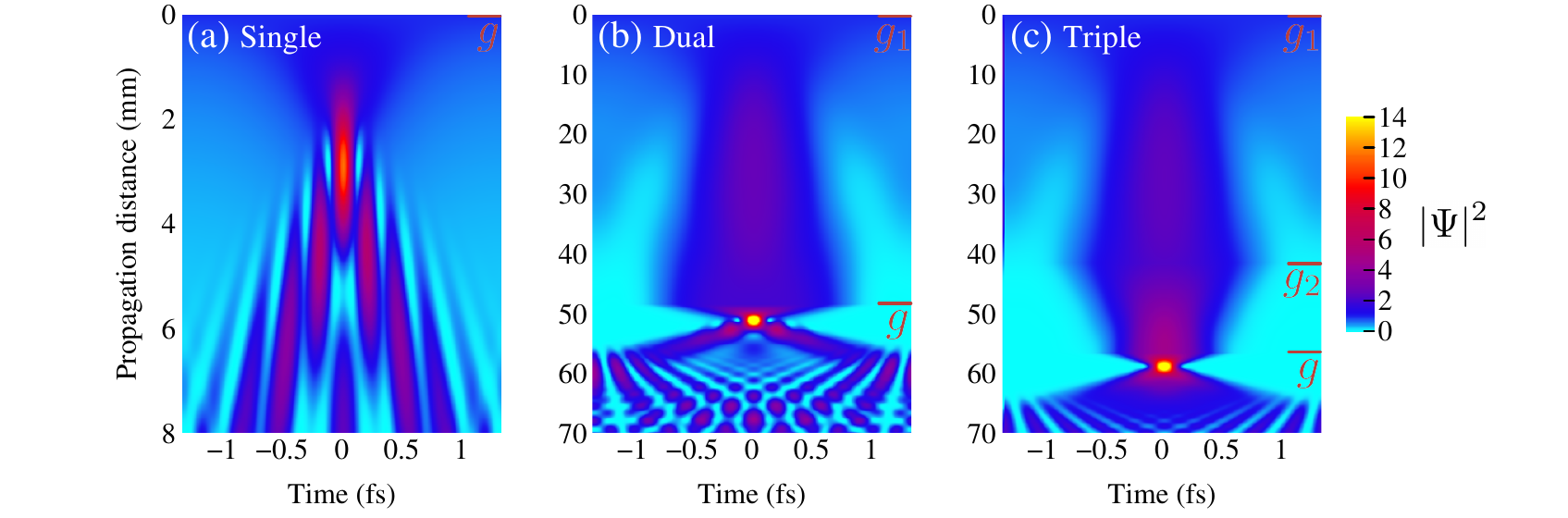}
\caption{Evolution of the probability density with propagation distance after interaction with 800-nm light. (a) Single interaction with a main compression strength $g=4$. (b) Numerically optimized dual interaction with a weak precompression strength $g_1=0.426$, drift distance $d_1=0.242 l_T$, and the same main compression strength $g=4$. (c) Optimized triple interaction with strengths $g_1=0.386$, $g_2=0.519$, $g=4$, and drift distances $d_1=0.207l_T$, $d_2=0.077l_T$. Red: The positions of the temporal phase plates and the respective interaction strengths.}\label{density_plot}
\end{figure}
\end{center}

Figures 2(a) and 2(b) in the main text illustrate temporal focusing with light, showing the magnitude of the wavefunction at selected planes. Here, in Figs.~\ref{density_plot}(a)-\ref{density_plot}(c), we provide a more detailed comparison of the three different focusing schemes for the same set of focusing parameters as in Fig.~2, showing the entire evolution of the densities as a function of distance.

\newpage
\onecolumngrid
\twocolumngrid

\newpage
\bibliography{references}{}